\documentclass[aps,pra,reprint,superscriptaddress]{revtex4-1}

\usepackage{amsmath}
\usepackage{amssymb}
\usepackage{mathtools}
\usepackage{bm}
\usepackage{braket}

\usepackage{graphicx}
\usepackage{subcaption}
\usepackage{float}

\usepackage{booktabs}

\usepackage{tikz}
\usetikzlibrary{arrows.meta}

\usepackage{xcolor}
\usepackage[
  colorlinks=true,
  linkcolor=red,
  citecolor=blue,
  urlcolor=magenta
]{hyperref}

\begin{document}

\title{Programmable Doppler real-space pattern formation in cold atoms}

\author{Yijun Tang}
\affiliation{
Cavendish Laboratory, University of Cambridge,
JJ Thomson Avenue, Cambridge CB3 0US, United Kingdom
}

\begin{abstract}
Red-detuned Doppler cooling decelerates atomic motion toward zero velocity,
whereas blue-detuned light produces acceleration that separates a cold cloud into finite-velocity packets. Here
we combine these complementary dynamics in a programmable split-stop
protocol for real-space pattern formation. During each stage, a
blue-detuned pulse splits existing packets and drives them apart, while
a subsequent red-detuned pulse returns their center-of-mass velocities
close to zero, results in spatially resolved stationary packets. Repeating this
process recursively multiplies the number of packets
in real space. We model the dynamics using stochastic photon-jump
simulations that include absorption and spontaneous-emission recoil in
one and two dimensions. Using experimentally realistic parameters for
the $689~\mathrm{nm}$ ${}^{1}S_{0}\rightarrow{}^{3}P_{1}$ transition of
${}^{88}\mathrm{Sr}$, we demonstrate an 8-packet one-dimensional
array and a 64-packet two-dimensional square array.

\end{abstract}

\maketitle

\section{Introduction}

A momentum-space crystal in cold atom is an atomic distribution in which the population occupies different velocity class, forming a discrete crystal like structure in momentum space. Its formation can be
understood from the Doppler force produced by a pair of counter-propagating
blue-detuned beams. For an atom moving along one beam, the
co-propagating light is Doppler shifted towards resonance, whereas the
counter-propagating light is shifted further from resonance. The
resulting imbalance in the scattering rates accelerates the atom along
its initial direction of motion. Consequently, an initially cold
ensemble separates into two oppositely moving subclouds whose
velocities become bunched around characteristic finite values. Therefore, the blue-detuned acceleration cause velocity bunching that
generates discrete momentum packets arranged according to the geometry
of the applied laser beams.

Momentum-space crystal in cold atom were first observed using narrow-line cooling
of strontium~\cite{loftus2004narrowprl,loftus2004narrowpra}. In
three-dimensional molasses, the momentum packets and
spatial distributions resembling the lattice points of cubic or
face-centered-cubic crystals, with their formations
controlled by the detuning, saturation, and interaction
time~\cite{loftus2004narrowpra}. Subsequent work revisited this
phenomenon also in strontium samples~\cite{stellmer2013degenerate}. More recent
experiments demonstrated momentum-space crystals in
fermionic $^{87}$Sr and compared their formation on narrow- and
broad-line transitions~\cite{han2019momentum}.

\begin{figure}[ht]
\centering
\includegraphics[width=\columnwidth]{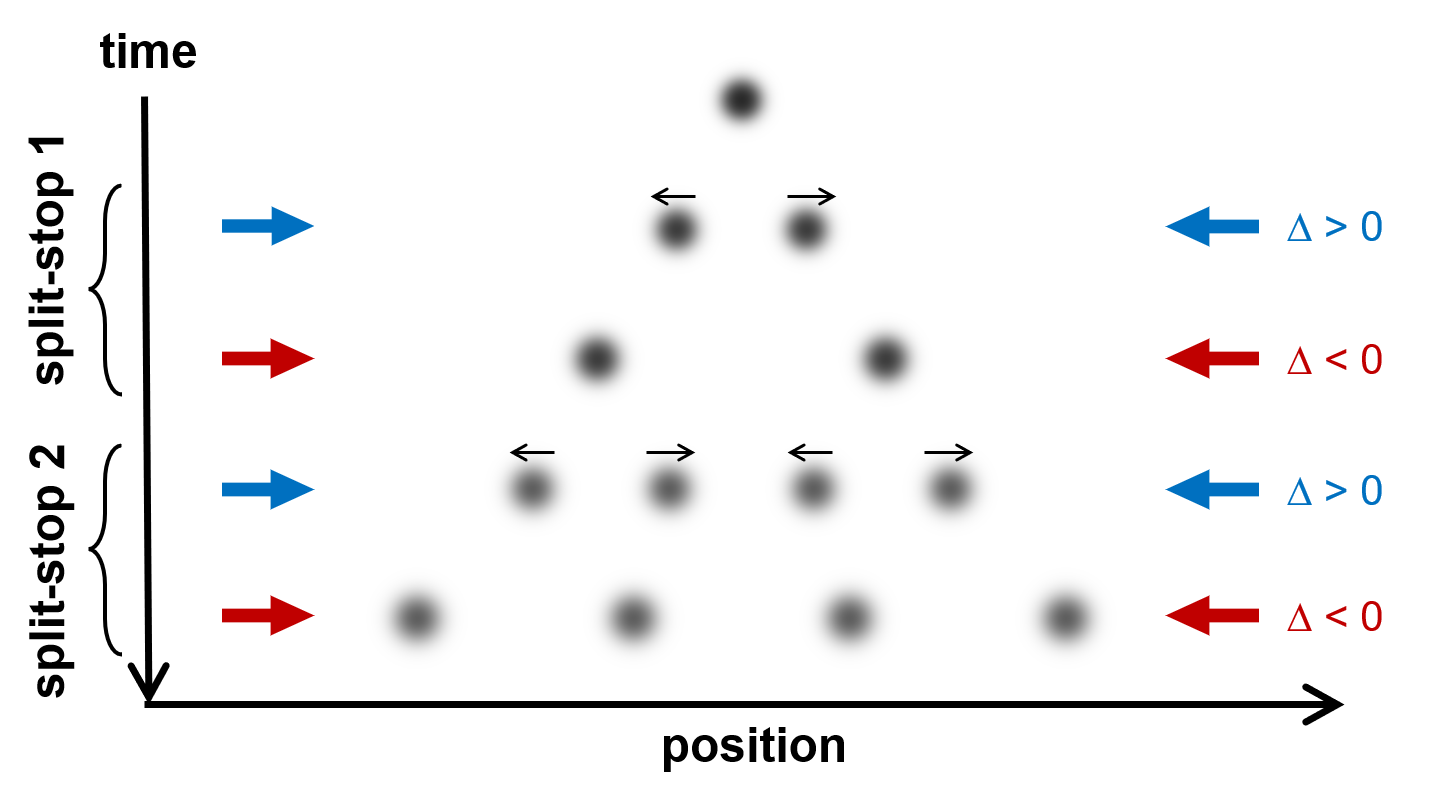}
\caption{
Schematic of pattern formation with alternating blue and red detuning. After two split-stop, the initial cloud splits into 4 subpackets. Arrows on the subpackets indicate their center-of-mass moving direction.
}
\label{fig:4_packet_example}
\end{figure}

Building on this blue-detuned splitting mechanism, we introduce the
split-stop protocol illustrated in
Fig.~\ref{fig:4_packet_example}. Starting from an initially cold atomic cloud, a blue-detuned pulse first divides the
cloud into two oppositely moving subpackets. Rather than retaining
these finite-velocity packets, a subsequent red-detuned pulse decelerates
them and brings their center-of-mass velocities close to zero \cite{Chu1985}. The detuning is then
returned to the blue side, causing each nearly stationary subpackets to
split again, after which another red-detuned pulse stops the newly formed
packets. Ideally, each split-stop cycle driven by one pair of
counter-propagating beams doubles the number of spatially resolved
packets along that axis. Each split-stop cycle drives an expansion and subsequent contraction of
the packet distribution in momentum space, repeatedly returning the
subpackets' center of mass velocity close to zero without forming a persistent
momentum-space crystal. Each such momentum-space ``breath'' multiplies
the number of spatially separated subpackets, thereby constructing a
progressively extended real-space pattern. By
programming the pulse durations, detunings, and saturation parameters,
we demonstrate the formation of extended one and two-dimensional
cold-atom packet arrays.

\section{1D Pattern Formation}

\subsection{model}

We investigate the 1D atomic motion under counter-propagating laser beams. The scattering rate of a two-level atom experiencing a single beam traveling in $\pm \hat{x}$ (subscript with $\pm$) is given by 

\begin{equation}
    R_{\pm}(v,t)
    =
    \frac{\Gamma}{2}
    \frac{q_{\pm}(v,t)}
    {1+q_{+}(v,t)+q_{-}(v,t)}.
    \label{eq:shared_scattering_rates}
\end{equation}

where $q_{\pm}(v,t)$ is detuning-weighted saturation parameters 
\begin{equation}
    q_{\pm}(v,t)
    =
    \frac{s(t)}
    {1+4\delta_{\pm}^{2}(v,t)/\Gamma^{2}},
    \qquad
    \delta_{\pm}(v,t)=\Delta(t)\mp kv,
    \label{eq:weighted_saturations}
\end{equation}
The $\pm$ sign is beam direction $s(t)=I/I_s$ is the single-beam saturation parameter with $I_s$ being saturation intensity, $\Gamma$ is natural linewidth, $\delta_{\pm}(v,t)$ is the detuning with $\Delta(t)$ being
the time-dependent laser detuning and $kv$ being the Doppler shift. To obtain eq.\ref{eq:shared_scattering_rates} and
\ref{eq:weighted_saturations}, we employ an incoherent two-level
rate-equation description inspired from literature \cite{minogin1985resonance,podlecki2017radiation,walker2026high}. The counter-propagating beams are assumed to
be mutually incoherent. The
internal atomic state is assumed to follow the instantaneous optical
steady state adiabatically. To simplify our study, the magnetic field and gravity effect are neglected. 

The simulation is implemented using a stochastic photon-jump process.
At each timestep $\Delta t$, the total effective scattering rate is
\begin{equation}
    R_{\rm sc}(v,t)=R_{+}(v,t)+R_{-}(v,t).
\end{equation}
For each atom, the number of complete scattering cycles occurring
during that timestep is sampled from a Poisson distribution. Thus, a timestep may contain zero, one, or multiple scattering cycles.
The timestep is chosen sufficiently small that the parameters and atomic dynamics changes are small. Each sampled scattering cycle consists of absorption from
one of the two laser beams followed by spontaneous emission.
Each effective scattering event is represented as a complete scattering cycle consisting of an absorption from one of the two laser beams followed by spontaneous emission. If a scattering event occurs, the probability that absorption is from $\pm\hat{x}$ beam is
\begin{equation}
P_{\pm}
=
\frac{R_{\pm}}
{R_{+}+R_{-}}
=
\frac{q_{\pm}}
{q_{+}+q_{-}},
\label{eq:beam_selection_probability}
\end{equation}
A second uniformly distributed random number is therefore used to select the absorbing beam. Absorption from the $+\hat{x}$ beam gives the atom a momentum recoil $+\hbar k$ whereas absorption from the $-\hat{x}$ beam gives $-\hbar k$.

The probabilities in Eq.~\eqref{eq:beam_selection_probability} describe two limiting regimes: effective single-beam addressing and simultaneous two-beam addressing. At sufficiently large atomic velocities, the Doppler shift brings one beam much closer to resonance while shifting the counter-propagating beam further away from resonance. For example, when $q_{+}\gg q_{-}$, one obtains $P_{+}\rightarrow1$, and the dynamics approach the single-beam-addressing limit. In contrast, near $v\simeq0$ for two beams of equal intensity and detuning, $q_{+}\simeq q_{-}$ and therefore $P_{+}\simeq P_{-}\simeq1/2$, so that both beams contribute comparably to the atomic dynamics. Large saturation parameters broaden the detuning range over which $q_{+}$ and $q_{-}$ are comparable, allowing simultaneous two-beam addressing to happen over a wider range of atomic velocities. 

Following the absorption recoil, the spontaneous-emission recoil is sampled independently. In the strictly one-dimensional recoil model, the atom receives a spontaneous-emission recoil of either $+\hbar k$ or $-\hbar k$, each with probability $1/2$. Since the absorption recoil is also $\pm\hbar k$, the total momentum change associated with a single effective scattering cycle can therefore take the values $-2\hbar k$, $0$, or $+2\hbar k$.

We then evolve the trajectories ${x_i,v_i}$ of an ensemble of atoms to the target time, with the saturation parameter $s(t)$ and detuning $\Delta(t)$ varied according to the designed pulse sequence.

\subsection{Pattern formation}

% Times t_j are measured in ms.
% delta(t) = Delta(t)/Gamma.
% s(t) is the saturation parameter of each counterpropagating beam.

We first demonstrate the protocol by generating a finite one-dimensional
array of eight spatially resolved atomic packets. 

The laser protocol is specified by two time-dependent control
parameters: the normalized detuning
$\delta(t)=\Delta(t)/\Gamma$ and the single-beam saturation parameter
$s(t)$, which is identical for the two counterpropagating beams.
Appropriate choices of these controls produce either blue-detuned
packet splitting or red-detuned packet stopping. We construct the protocol from $n$ successive time segments with
boundaries $t_0<t_1<\cdots<t_n.$ 
Within each segment, the parameters are  linearly 
swept between start and end points of the segment. The endpoint values
of adjacent segments need not coincide. For segment $j=1,\ldots,n$, we define
the normalized local time

\begin{equation}
    \tau_j(t)
    =
    \frac{t-t_{j-1}}{t_j-t_{j-1}},
    \qquad
    t_{j-1}\leq t<t_j .
    \label{eq:local_segment_time}
\end{equation}
The complete control sequence is then
\begin{subequations}
\label{eq:pulse_sequence}
\begin{align}
\delta(t)
&=
\begin{cases}
0, & t<0,\\[2pt]
d_j^-+\left(d_j^+-d_j^-\right)\tau_j(t),
& t_{j-1}\leq t<t_j,\\[2pt]
0, & t\geq t_{n},
\end{cases}
\label{eq:pulse_detuning}
\\[6pt]
s(t)
&=
\begin{cases}
0, & t<0,\\
s_j, & t_{j-1}\leq t<t_j,\\
0, & t\geq t_{n}.
\end{cases}
\label{eq:pulse_saturation}
\end{align}
\end{subequations}
Here, $d_j^-$ and $d_j^+$ are the initial and final normalized
detunings of segment $j$, respectively, and $s_j$ is the saturation
parameter of the pair of counterpropagating beam. 

We initialize a one-dimensional ensemble of $10^{5}$ $^{88}\mathrm{Sr}$ atoms with a Gaussian spatial distribution of width $\sigma_x=150~\mu\mathrm{m}$ and a temperature of $T=800~\mathrm{nK}$. These parameters are representative of experimentally demonstrated narrow-line red-MOT conditions for $^{88}\mathrm{Sr}$ using the $689~\mathrm{nm}$ $^{1}S_{0}\rightarrow{}^{3}P_{1}$ transition, which is also used throughout our simulations~\cite{pasatembou2024progress, katori1999magneto}. The atomic trajectories are propagated using a fixed timestep of $dt=5~\mu\mathrm{s}$.

\begin{figure*}{h}
    \centering
    \includegraphics[width=0.9\textwidth]{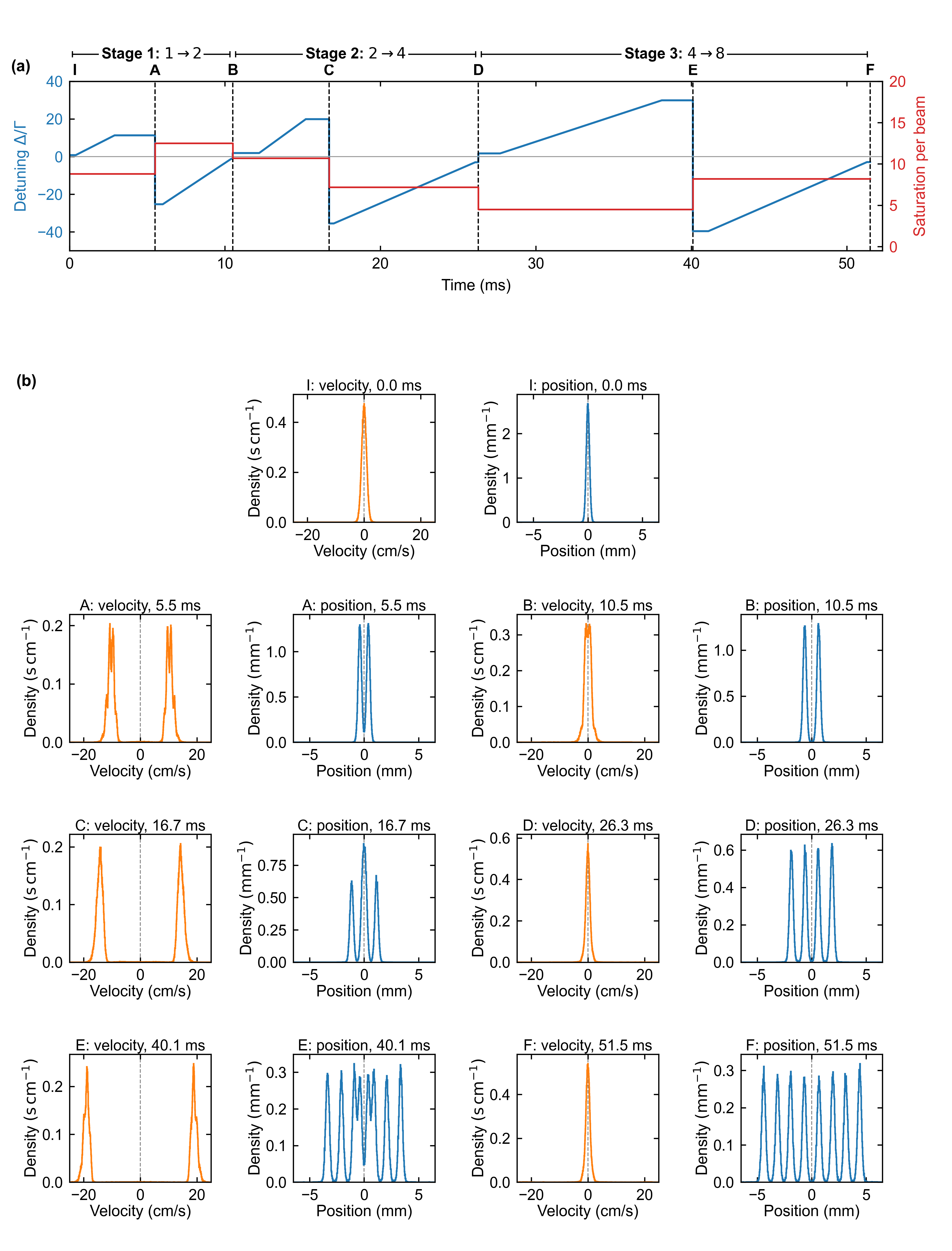}
    \caption{
    Hierarchical formation of an eight-packet one-dimensional real-space
    distribution using the split-stop protocol.
    (a) Time-dependent normalized detuning $\Delta(t)/\Gamma$ (blue) and
    single-beam saturation parameter $s(t)$ (red). The three stages produce
    the successive $1\rightarrow2$, $2\rightarrow4$, and
    $4\rightarrow8$ packet configurations. Dashed lines indicate the times I--F.
    (b) Corresponding velocity-space (orange) and real-space (blue)
    probability densities. I is initial cloud distribution. A, C, and E shows 2,4 and 8 subpackets at the end of blue-detuned pulses. Subsequent red-detuned pulses reduce
    their mean velocities to rest at B, D, and F, respectively. Detailed sequence parameters are in Table.\ref{tab:pulse_sequence} in Appendix.\ref{app:1d_sequence}.
    }
    \label{fig:split_stop_1d}
\end{figure*}

As shown in Fig.~\ref{fig:split_stop_1d}, the protocol comprises three
stages that successively increase the number of packets from
$1\rightarrow2$, $2\rightarrow4$, and $4\rightarrow8$.
Each stage consists of a blue-detuned splitting subsequence followed by a
red-detuned stopping subsequence. Both subsequences employ a
hold--ramp--hold structure: the detuning is first held constant, then
swept linearly, and finally held at its terminal value. The saturation
parameter remains constant within each subsequence but may change between
the splitting and stopping subsequences and between successive stages.

At $t=0$, the ensemble consists of a single Gaussian cloud. Stage~1
begins with a $0.4~\mathrm{ms}$ blue-detuned seed pulse at
$\Delta=0.8\Gamma$, which preferentially accelerates the slower atoms
in opposite directions. As their Doppler shifts increase, the detuning
is swept linearly to $11.3\Gamma$ over $2.5~\mathrm{ms}$, thereby
tracking the accelerating atoms and driving the two subpackets to higher
velocities. The detuning is then held at $11.3\Gamma$ for
$2.6~\mathrm{ms}$. This terminal hold bunches the atoms around their
final velocities while allowing the two subpackets additional time to
separate spatially. The velocity separation is largest at the end of
this blue-detuned hold, as shown by snapshot~A in
Fig.~\ref{fig:split_stop_1d}(b).

The stopping subsequence begins by jumping the detuning to
$-25.3\Gamma$, which is held for $0.5~\mathrm{ms}$ to address the
rapidly moving atoms. The detuning is subsequently swept towards
resonance, reaching $-1.2\Gamma$ after $4.4~\mathrm{ms}$. This sweep
follows the decreasing Doppler shift of the decelerating atoms and
maintains the stopping force as their velocities approach zero. Because
the packets are nearly stationary at the end of the ramp, only a short
$0.1~\mathrm{ms}$ terminal hold is required. As illustrated by
snapshot~B, the velocity distribution is once again centred near zero,
while two well-separated spatial subpackets have formed.

The same split-stop principle is applied in stages~2 and~3 to
produce four and eight packets, respectively. The saturation parameters
of the splitting and stopping subsequences are numerically optimized to
maximize the contrast between neighboring packets while maintaining
narrow packet widths. The later stages
require longer sweeps and holds because the ensemble already spans
several millimeters, so the oppositely moving groups must travel farther
to become fully separated. At the end of the complete sequence,
$t=51.5~\mathrm{ms}$, the ensemble forms eight well-resolved spatial
packets. The variance of the final velocity distribution corresponds to
an effective one-dimensional kinetic temperature of approximately
$1.2~\mu\mathrm{K}$.

\section{2D Pattern Formation}

\subsection{model}

The two-dimensional model is a direct vector extension of the 1D model built above. We consider four laser beams propagating
along
$\mathbf n_i\in\{+\hat{\mathbf x},-\hat{\mathbf x},
+\hat{\mathbf y},-\hat{\mathbf y}\}$.
The two counter-propagating pairs may have independently programmed
detunings $\Delta_x(t)$ and $\Delta_y(t)$ and single-beam saturation
parameters $s_x(t)$ and $s_y(t)$. For beam $i$, we define
\begin{equation}
q_i(\mathbf v,t)
=
\frac{s_i(t)}
{1+4\left[\Delta_i(t)-k\mathbf n_i\cdot\mathbf v\right]^2/\Gamma^2},
\end{equation}
where $(\Delta_i,s_i)=(\Delta_x,s_x)$ for the $\pm\hat{\mathbf x}$
beams and $(\Delta_i,s_i)=(\Delta_y,s_y)$ for the
$\pm\hat{\mathbf y}$ beams. Scattering rate from beam $i$ is
\begin{equation}
R_i(\mathbf v,t)
=
\frac{\Gamma}{2}
\frac{q_i(\mathbf v,t)}
{1+\sum_j q_j(\mathbf v,t)},
\label{eq:2d_scattering_rates}
\end{equation}

Total scattering rate is $R_{\mathrm{sc}}=\sum_i R_i$. At each timestep, the number of scattering cycles experienced by each atom is randomly sampled from a Poisson distribution with mean \(R_{\mathrm{sc}}\Delta t\).
For each cycle, the absorbing beam is selected with probability
\begin{equation}
P_i=\frac{R_i}{R_{\mathrm{sc}}}
=\frac{q_i}{\sum_jq_j},
\end{equation}
and contributes an atomic momentum kick $\hbar k\mathbf n_i$.
Spontaneous emission is sampled isotropically in three dimensions by
drawing $\mu=\cos\theta$ uniformly from $[-1,1]$ and $\phi$ uniformly
from $[0,2\pi]$, giving
\begin{equation}
\mathbf n_{\mathrm{sp}}
=
\left(
\sqrt{1-\mu^2}\cos\phi,\,
\sqrt{1-\mu^2}\sin\phi,\,
\mu
\right).
\end{equation}
The in-plane momentum change for one scattering cycle is therefore
\begin{equation}
\Delta\mathbf p_{xy}
=
\hbar k
\left(
\mathbf n_i-\mathbf n_{\mathrm{sp}}
\right)_{xy}.
\label{eq:2d_recoil}
\end{equation}

\subsection{Pattern formation}

We will present a demonstration the formation
of an $8\times8$ square array of 64 spatially resolved atomic packets.

We consider simultaneous addressing by the two counterpropagating beam
pairs along $x$ and $y$, with identical detuning and saturation parameters,
\begin{equation}
    \Delta_x(t)=\Delta_y(t)\equiv\Delta(t),
    \qquad
    s_x(t)=s_y(t)\equiv s(t).
\end{equation}
Both beam pairs are therefore governed by the same sequence
parameterization introduced in
eq.~\ref{eq:local_segment_time},
\ref{eq:pulse_detuning}, and
\ref{eq:pulse_saturation}.

\begin{figure*}
    \centering
    \includegraphics[width=0.9\textwidth]{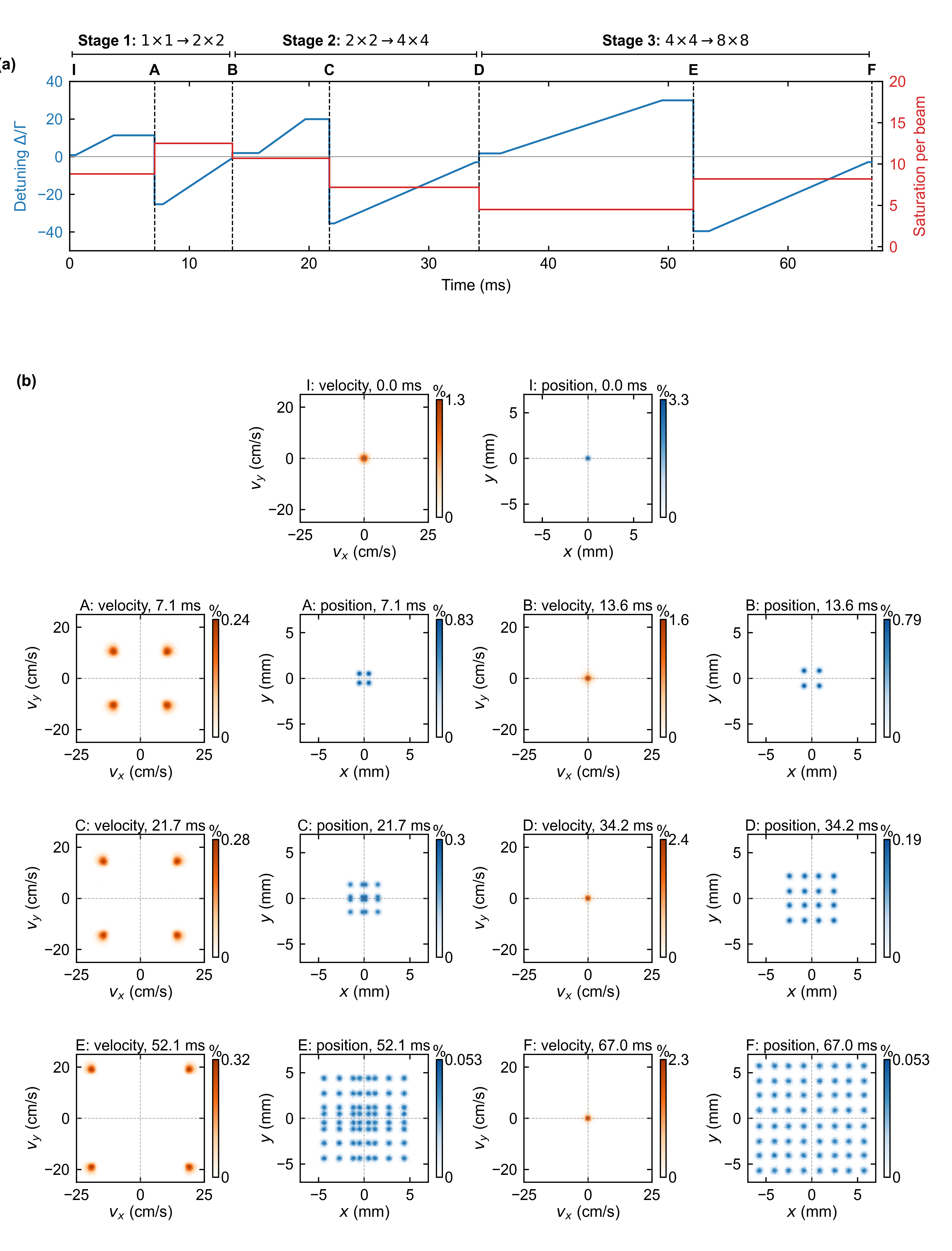}
    \caption{
Hierarchical formation of an $8\times8$ two-dimensional real-space
packet array using simultaneous two-axis split-stop control.
(a) Normalized detuning $\Delta(t)/\Gamma$ (blue) and single-beam
saturation $s(t)$ (red), applied identically along $x$ and $y$.
The three stages produce the successive
$1\times1\rightarrow2\times2\rightarrow4\times4\rightarrow8\times8$
configurations. Dashed lines mark times A--F.
(b) Corresponding velocity-space (orange) and real-space (blue)
probability densities, including the initial distributions I.
Blue-detuned pulses produce four-quadrant velocity splitting at A, C,
and E; red-detuned pulses return the packets' velocity toward rest at B, D, and F. The independently scaled color bars show the percentage of the full atomic ensemble contained in each fixed two-dimensional histogram bin. Detailed sequence parameters are in Table.\ref{tab:2d_pulse_sequence} in Appendix.\ref{app:2d_sequence}.
}
    \label{fig:split_stop_2d}
\end{figure*}

During a blue-detuned splitting subsequence, the Doppler force amplifies atoms according to the initial signs of the velocity
components. For example, an atom with $v_x>0$ preferentially scatters
from the $+\hat{\mathbf{x}}$ beam, further increasing $v_x$, whereas an
atom with $v_x<0$ is preferentially driven in the opposite direction.
An analogous process occurs along $y$. The initial ensemble consequently
separates into four velocity groups associated with the quadrants
\begin{equation}
    (v_x,v_y)\sim(+,+),\;(+,-),\;(-,+),\;(-,-).
\end{equation}
Although each scattering event involves absorption from only one of the
four beams, the accumulation of many events transfers momentum along
both axes and drives the four groups toward the diagonal directions.

The subsequent red-detuned subsequence reverses this to produce a standard molasses.
Along each axis, atoms preferentially scatter from the beam opposing
their motion, bringing the four velocity groups back toward
$\mathbf v=\mathbf 0$, leaving four approximately stationary and spatially separated packets.
Each complete split-stop stage therefore increases the number of stationary packets by a factor of four.

Applying this operation recursively produces
the hierarchy
\begin{equation}
    1\times1
    \longrightarrow 2\times2
    \longrightarrow 4\times4
    \longrightarrow 8\times8,
\end{equation}
corresponding to $1$, $4$, $16$, and finally $64$ spatial packets as shown in Fig.\ref{fig:split_stop_2d}.
Similar to the 1D protocol, the saturation parameters of the blue-detuned
splitting and red-detuned stopping subsequences are optimized
numerically within the chosen experimental bounds to maximize packet
separation and contrast while limiting packet broadening. At the end of the complete sequence,
$t=67.0~\mathrm{ms}$, the ensemble has an effective two-dimensional kinetic temperature of approximately
$0.8~\mu\mathrm{K}$.

An alternative strategy is to control the $x$ and $y$ beam pairs
independently, producing the sequential hierarchy
$1\rightarrow2\rightarrow4\rightarrow8\rightarrow16\rightarrow32
\rightarrow64$. This axis-by-axis protocol provides greater control
over the pattern geometry at the cost of a
longer formation time. In the example considered here, producing 64
packets requires $113.2~\mathrm{ms}$, compared with $67.0~\mathrm{ms}$
for simultaneous $xy$ control. Details of the axis-by-axis protocol
are provided in Appendix.\ref{app:2d_sequence_axis}.

\section{Conclusion}
\label{sec:conclusion}

We introduce a split-stop protocol for recursively multiplying spatially resolved cold-atom packets. Using a stochastic photon-jump model in one and two dimensions, we simulate experimentally realistic implementations on the \(689~\mathrm{nm}\) transition of \(^{88}\mathrm{Sr}\), demonstrating an eight-packet 1D array and a 64-packet 2D square array.

Several directions naturally follow from this work. First, alternative beam configurations and addressing sequences could
generate a wider range of packet geometries. Although the resulting
patterns are neither single-atom resolved nor held in confining
potentials, this approach may provide a comparatively coarse and simple
method for preparing structured atomic density distributions without
spatially patterned light.\cite{henderson2009painting,nogrette2014single,
barredo2016assembler,endres2016assembly}. Second,
optimal-control methods could be used to design nonlinear detuning and
intensity profiles that reduce the total sequence duration while maximizing
packet contrast and minimizing packet width. Finally, although our example
uses the narrow $689~\mathrm{nm}$ transition of strontium, the protocol could
be extended to other atomic species, such as Ytterbium. 

\clearpage
\onecolumngrid
\appendix

\section{1D sequence parameters}
\label{app:1d_sequence}

The control parameters used to generate
Fig.~\ref{fig:split_stop_1d} are listed in
Table~\ref{tab:pulse_sequence}.

\begin{table*}[h]
\caption{
Control parameters for the three split-stop stages producing the
successive $1\rightarrow2$, $2\rightarrow4$, and
$4\rightarrow8$ packet configurations.
All times are in milliseconds.
The quantities $d_j^-$ and $d_j^+$ are the normalized detunings at
the beginning and end of segment $j$, respectively, and $s_j$ is the
single-beam saturation parameter, identical for the two
counter-propagating beams.
}
\label{tab:pulse_sequence}
\centering
\begingroup
\small
\setlength{\tabcolsep}{2.2pt}
\renewcommand{\arraystretch}{1.05}

\begin{ruledtabular}
\begin{tabular}{
crrrrr@{\hspace{1.2em}}
crrrrr@{\hspace{1.2em}}
crrrrr
}
\multicolumn{6}{c}{Stage 1: $1\rightarrow2$}
&
\multicolumn{6}{c}{Stage 2: $2\rightarrow4$}
&
\multicolumn{6}{c}{Stage 3: $4\rightarrow8$}
\\
$j$ & $t_{j-1}$ & $t_j$ & $d_j^-$ & $d_j^+$ & $s_j$
&
$j$ & $t_{j-1}$ & $t_j$ & $d_j^-$ & $d_j^+$ & $s_j$
&
$j$ & $t_{j-1}$ & $t_j$ & $d_j^-$ & $d_j^+$ & $s_j$
\\
\hline
1 & 0.0 & 0.4 & 0.8 & 0.8 & 8.8
&
7 & 10.5 & 12.2 & 1.9 & 1.9 & 10.7
&
13 & 26.3 & 27.7 & 1.7 & 1.7 & 4.5
\\
2 & 0.4 & 2.9 & 0.8 & 11.3 & 8.8
&
8 & 12.2 & 15.2 & 1.9 & 19.9 & 10.7
&
14 & 27.7 & 38.1 & 1.7 & 29.9 & 4.5
\\
3 & 2.9 & 5.5 & 11.3 & 11.3 & 8.8
&
9 & 15.2 & 16.7 & 19.9 & 19.9 & 10.7
&
15 & 38.1 & 40.1 & 29.9 & 29.9 & 4.5
\\
4 & 5.5 & 6.0 & -25.3 & -25.3 & 12.5
&
10 & 16.7 & 17.0 & -35.5 & -35.5 & 7.2
&
16 & 40.1 & 41.1 & -39.6 & -39.6 & 8.2
\\
5 & 6.0 & 10.4 & -25.3 & -1.2 & 12.5
&
11 & 17.0 & 26.1 & -35.5 & -3.0 & 7.2
&
17 & 41.1 & 51.3 & -39.6 & -2.9 & 8.2
\\
6 & 10.4 & 10.5 & -1.2 & -1.2 & 12.5
&
12 & 26.1 & 26.3 & -3.0 & -3.0 & 7.2
&
18 & 51.3 & 51.5 & -2.9 & -2.9 & 8.2
\\
\end{tabular}
\end{ruledtabular}

\endgroup
\end{table*}

\section{2D sequence parameters}
\label{app:2d_sequence}

The control parameters used to generate
Fig.~\ref{fig:split_stop_2d} are listed in
Table~\ref{tab:2d_pulse_sequence}. The same controls are applied
simultaneously to the counterpropagating beam pairs along $x$ and $y$.

\begin{table*}[h]
\caption{
Control parameters for the three simultaneous two-axis split-stop
stages producing the successive
$1\times1\rightarrow2\times2$,
$2\times2\rightarrow4\times4$, and
$4\times4\rightarrow8\times8$ packet configurations.
All times are in milliseconds. The quantities $d_j^-$ and $d_j^+$ are
the normalized detunings at the beginning and end of segment $j$,
respectively, and $s_j$ is the single-beam saturation parameter.
Identical values are applied to the $x$ and $y$ beam pairs.
}
\label{tab:2d_pulse_sequence}
\centering
\begingroup
\small
\setlength{\tabcolsep}{2.2pt}
\renewcommand{\arraystretch}{1.05}

\begin{ruledtabular}
\begin{tabular}{
crrrrr@{\hspace{1.2em}}
crrrrr@{\hspace{1.2em}}
crrrrr
}
\multicolumn{6}{c}{Stage 1: $1\times1\rightarrow2\times2$}
&
\multicolumn{6}{c}{Stage 2: $2\times2\rightarrow4\times4$}
&
\multicolumn{6}{c}{Stage 3: $4\times4\rightarrow8\times8$}
\\
$j$ & $t_{j-1}$ & $t_j$ & $d_j^-$ & $d_j^+$ & $s_j$
&
$j$ & $t_{j-1}$ & $t_j$ & $d_j^-$ & $d_j^+$ & $s_j$
&
$j$ & $t_{j-1}$ & $t_j$ & $d_j^-$ & $d_j^+$ & $s_j$
\\
\hline
1 & 0.0 & 0.5 & 0.8 & 0.8 & 8.8
&
7 & 13.6 & 15.8 & 1.9 & 1.9 & 10.7
&
13 & 34.2 & 36.0 & 1.7 & 1.7 & 4.5
\\
2 & 0.5 & 3.7 & 0.8 & 11.3 & 8.8
&
8 & 15.8 & 19.7 & 1.9 & 19.9 & 10.7
&
14 & 36.0 & 49.5 & 1.7 & 29.9 & 4.5
\\
3 & 3.7 & 7.1 & 11.3 & 11.3 & 8.8
&
9 & 19.7 & 21.7 & 19.9 & 19.9 & 10.7
&
15 & 49.5 & 52.1 & 29.9 & 29.9 & 4.5
\\
4 & 7.1 & 7.8 & -25.3 & -25.3 & 12.5
&
10 & 21.7 & 22.1 & -35.5 & -35.5 & 7.2
&
16 & 52.1 & 53.4 & -39.6 & -39.6 & 8.2
\\
5 & 7.8 & 13.5 & -25.3 & -1.2 & 12.5
&
11 & 22.1 & 33.9 & -35.5 & -3.0 & 7.2
&
17 & 53.4 & 66.7 & -39.6 & -2.9 & 8.2
\\
6 & 13.5 & 13.6 & -1.2 & -1.2 & 12.5
&
12 & 33.9 & 34.2 & -3.0 & -3.0 & 7.2
&
18 & 66.7 & 67.0 & -2.9 & -2.9 & 8.2
\\
\end{tabular}
\end{ruledtabular}

\endgroup
\end{table*}

\section{2D axis-by-axis addressing}
\label{app:2d_sequence_axis}

\begin{figure*}
    \centering
    \includegraphics[width=0.9\textwidth]{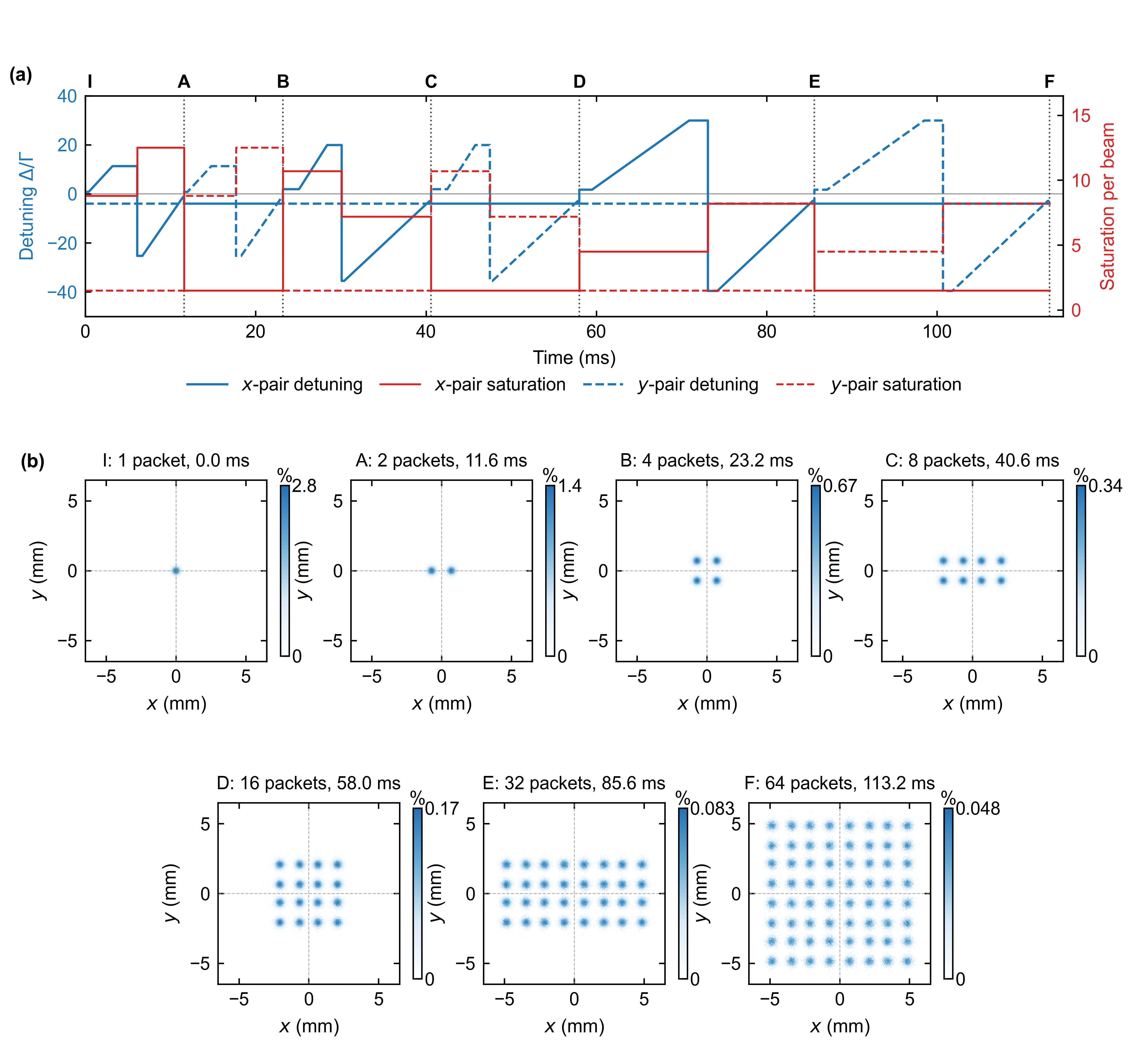}
    \caption{
Hierarchical formation of an $8\times8$ two-dimensional real-space
packet array using independent axis-by-axis split-and-stop control.
(a) Independently programmed normalized detunings and single-beam
saturations for the $x$ and $y$ beam pairs. 
(b) Corresponding real-space probability distributions. Alternating
control of the two axes successively produces
$1\rightarrow2\rightarrow4\rightarrow8\rightarrow16\rightarrow32
\rightarrow64$ spatial packets, culminating in an $8\times8$ array.
The independently scaled color bars show the percentage of the full
atomic ensemble contained in each fixed two-dimensional histogram bin.
}
    \label{fig:split_stop_2d_axis}
\end{figure*}

An alternative strategy is to address the $x$ and $y$ beam pairs
sequentially, as illustrated in Fig.~\ref{fig:split_stop_2d_axis}.
During each split-stop operation, one beam pair actively divides
the cloud along its axis, while the orthogonal pair is held at a
red-detuned molasses that suppresses
expansion in the other direction. Alternating the addressed axis
therefore produces the hierarchy $1\rightarrow2\rightarrow4\rightarrow8 \rightarrow16\rightarrow32\rightarrow64$ spatial packets. This axis-by-axis protocol provides independent control over the two
directions and therefore offers greater flexibility in determining the pattern geometry. The trade-off
is a longer sequence: each operation doubles the number of packets,
whereas simultaneous addressing of both axes can increase the packet
number by a factor of four in a single split-stop stage.

The axis-by-axis protocol uses the same piecewise-linear
segment definitions in
eq.~\ref{eq:local_segment_time}--\ref{eq:pulse_saturation}, but
the two counterpropagating beam pairs are programmed independently.
For each axis $\alpha\in\{x,y\}$, we divide the complete protocol into
$n_\alpha=21$ contiguous segments with boundaries
$t_{\alpha,0}<\cdots<t_{\alpha,n_\alpha}$ and define $\tau_{\alpha j}(t)
    =
    (t-t_{\alpha,j-1})/
         (t_{\alpha,j}-t_{\alpha,j-1}),$
    $t_{\alpha,j-1}\leq t<t_{\alpha,j}.$

The controls applied to beam pair $\alpha$ are
\begin{subequations}
\begin{align}
\delta_\alpha(t)
&=
\begin{cases}
0, & t<0,\\[2pt]
d_{\alpha j}^{-}
+\left(d_{\alpha j}^{+}-d_{\alpha j}^{-}\right)
 \tau_{\alpha j}(t),
& t_{\alpha,j-1}\leq t<t_{\alpha,j},\\[2pt]
0, & t\geq t_{\alpha,n_\alpha},
\end{cases}
\\[6pt]
s_\alpha(t)
&=
\begin{cases}
0, & t<0,\\
s_{\alpha j},
& t_{\alpha,j-1}\leq t<t_{\alpha,j},\\
0, & t\geq t_{\alpha,n_\alpha}.
\end{cases}
\end{align}
\end{subequations}
The red-detuned periods are included explicitly as segments with
$d_{\alpha j}^{-}=d_{\alpha j}^{+}=-4$ and
$s_{\alpha j}=1.5$. The complete $x$ and $y$ sequences are listed
in Tables~\ref{tab:axis_x_sequence} and
\ref{tab:axis_y_sequence}, respectively.

\begin{table*}[t]
\caption{
Control parameters for the $x$ beam pair in the axis-by-axis
protocol. All times are in milliseconds.
}
\label{tab:axis_x_sequence}
\centering
\begingroup
\small
\setlength{\tabcolsep}{2.2pt}
\renewcommand{\arraystretch}{1.05}
\begin{ruledtabular}
\begin{tabular}{
crrrrr@{\hspace{1.2em}}
crrrrr@{\hspace{1.2em}}
crrrrr
}
\multicolumn{6}{c}{$1\rightarrow2\rightarrow4$}
&
\multicolumn{6}{c}{$4\rightarrow8\rightarrow16$}
&
\multicolumn{6}{c}{$16\rightarrow32\rightarrow64$}
\\
$j$ & $t_{j-1}$ & $t_j$ & $d_j^-$ & $d_j^+$ & $s_j$
&
$j$ & $t_{j-1}$ & $t_j$ & $d_j^-$ & $d_j^+$ & $s_j$
&
$j$ & $t_{j-1}$ & $t_j$ & $d_j^-$ & $d_j^+$ & $s_j$
\\
\hline
1 & 0.0 & 0.4 & 0.8 & 0.8 & 8.8
&
8 & 23.2 & 25.1 & 1.9 & 1.9 & 10.7
&
15 & 58.0 & 59.5 & 1.7 & 1.7 & 4.5
\\
2 & 0.4 & 3.2 & 0.8 & 11.3 & 8.8
&
9 & 25.1 & 28.4 & 1.9 & 19.9 & 10.7
&
16 & 59.5 & 70.9 & 1.7 & 29.9 & 4.5
\\
3 & 3.2 & 6.1 & 11.3 & 11.3 & 8.8
&
10 & 28.4 & 30.1 & 19.9 & 19.9 & 10.7
&
17 & 70.9 & 73.1 & 29.9 & 29.9 & 4.5
\\
4 & 6.1 & 6.7 & -25.3 & -25.3 & 12.5
&
11 & 30.1 & 30.4 & -35.5 & -35.5 & 7.2
&
18 & 73.1 & 74.2 & -39.6 & -39.6 & 8.2
\\
5 & 6.7 & 11.5 & -25.3 & -1.2 & 12.5
&
12 & 30.4 & 40.4 & -35.5 & -3.0 & 7.2
&
19 & 74.2 & 85.4 & -39.6 & -2.9 & 8.2
\\
6 & 11.5 & 11.6 & -1.2 & -1.2 & 12.5
&
13 & 40.4 & 40.6 & -3.0 & -3.0 & 7.2
&
20 & 85.4 & 85.6 & -2.9 & -2.9 & 8.2
\\
7 & 11.6 & 23.2 & -4.0 & -4.0 & 1.5
&
14 & 40.6 & 58.0 & -4.0 & -4.0 & 1.5
&
21 & 85.6 & 113.2 & -4.0 & -4.0 & 1.5
\\
\end{tabular}
\end{ruledtabular}
\endgroup
\end{table*}

\begin{table*}[t]
\caption{
Control parameters for the $y$ beam pair in the axis-by-axis
protocol. All times are in milliseconds.
}
\label{tab:axis_y_sequence}
\centering
\begingroup
\small
\setlength{\tabcolsep}{2.2pt}
\renewcommand{\arraystretch}{1.05}
\begin{ruledtabular}
\begin{tabular}{
crrrrr@{\hspace{1.2em}}
crrrrr@{\hspace{1.2em}}
crrrrr
}
\multicolumn{6}{c}{$1\rightarrow2\rightarrow4$}
&
\multicolumn{6}{c}{$4\rightarrow8\rightarrow16$}
&
\multicolumn{6}{c}{$16\rightarrow32\rightarrow64$}
\\
$j$ & $t_{j-1}$ & $t_j$ & $d_j^-$ & $d_j^+$ & $s_j$
&
$j$ & $t_{j-1}$ & $t_j$ & $d_j^-$ & $d_j^+$ & $s_j$
&
$j$ & $t_{j-1}$ & $t_j$ & $d_j^-$ & $d_j^+$ & $s_j$
\\
\hline
1 & 0.0 & 11.6 & -4.0 & -4.0 & 1.5
&
8 & 23.2 & 40.6 & -4.0 & -4.0 & 1.5
&
15 & 58.0 & 85.6 & -4.0 & -4.0 & 1.5
\\
2 & 11.6 & 12.0 & 0.8 & 0.8 & 8.8
&
9 & 40.6 & 42.5 & 1.9 & 1.9 & 10.7
&
16 & 85.6 & 87.1 & 1.7 & 1.7 & 4.5
\\
3 & 12.0 & 14.8 & 0.8 & 11.3 & 8.8
&
10 & 42.5 & 45.8 & 1.9 & 19.9 & 10.7
&
17 & 87.1 & 98.5 & 1.7 & 29.9 & 4.5
\\
4 & 14.8 & 17.7 & 11.3 & 11.3 & 8.8
&
11 & 45.8 & 47.5 & 19.9 & 19.9 & 10.7
&
18 & 98.5 & 100.7 & 29.9 & 29.9 & 4.5
\\
5 & 17.7 & 18.3 & -25.3 & -25.3 & 12.5
&
12 & 47.5 & 47.8 & -35.5 & -35.5 & 7.2
&
19 & 100.7 & 101.8 & -39.6 & -39.6 & 8.2
\\
6 & 18.3 & 23.1 & -25.3 & -1.2 & 12.5
&
13 & 47.8 & 57.8 & -35.5 & -3.0 & 7.2
&
20 & 101.8 & 113.0 & -39.6 & -2.9 & 8.2
\\
7 & 23.1 & 23.2 & -1.2 & -1.2 & 12.5
&
14 & 57.8 & 58.0 & -3.0 & -3.0 & 7.2
&
21 & 113.0 & 113.2 & -2.9 & -2.9 & 8.2
\\
\end{tabular}
\end{ruledtabular}
\endgroup
\end{table*}

\twocolumngrid
\bibliographystyle{apsrev4-1}
\bibliography{ref}

\end{document}